\documentclass[aps,prx,twocolumn,amsmath,amssymb,superscriptaddress,floatfix]{revtex4-2}

\usepackage{mathtools}
\usepackage{multirow}
\usepackage{xcolor}
\usepackage{bm}
\usepackage{amsfonts,amssymb,amsmath}
\usepackage{graphicx,dcolumn,bm,xcolor,braket,slashed}
\usepackage{times} 
\usepackage{comment}
\usepackage{array}
\usepackage{textcomp}
\usepackage[normalem]{ulem}
\usepackage{dsfont}
\usepackage{kotex}

\usepackage[title,toc,titletoc,page]{appendix}

\newcommand{\be}{\begin{eqnarray}}
\newcommand{\ee}{\end{eqnarray}}
\newcommand{\bbm}{\begin{bmatrix}}
\newcommand{\ebm}{\end{bmatrix}}
\newcommand{\bpm}{\begin{pmatrix}}
\newcommand{\epm}{\end{pmatrix}}

\begin{document}
\title{Orbital Embedding and the Physical Definition of Quantum Geometry}

\author{Chang-geun Oh}
\email{cg.oh.0404@gmail.com}
\affiliation{Department of Applied Physics, The University of Tokyo, Tokyo 113-8656, Japan}
\author{Shuichi Murakami}
\email{murakami@ap.t.u-tokyo.ac.jp}
\affiliation{Department of Applied Physics, The University of Tokyo, Tokyo 113-8656, Japan}
\affiliation{RIKEN Center for Emergent Matter Science (CEMS), 2-1 Hirosawa, Wako, Saitama, 351-0198, Japan}
\affiliation{
International Institute for Sustainability with Knotted Chiral Meta Matter (WPI-SKCM$^2$),Hiroshima University, 1-3-1 Kagamiyama, Higashi-Hiroshima, Hiroshima 739-8526, Japan}

             
\begin{abstract}
The Quantum Geometric Tensor, encompassing the quantum metric and Berry curvature, is a central concept in modern condensed matter physics. 
However, its standard calculation via $k$-derivatives of the Bloch projector conceals a fundamental ambiguity regarding the choice of unit-cell convention, specifically in the treatment of intra-cell orbital positions (i.e., with or without the orbital position $e^{ikx_\alpha}$).
We resolve this inconsistency by introducing a convention-independent physical QGT defined via a covariant derivative that explicitly incorporates the full position operator. We demonstrate that this formulation is uniquely mandated by the microscopic derivation of the physical current via the Peierls substitution. 
Notably, we uncover a leading-order failure in standard $k \cdot p$ effective theories for systems with bond-ordered gaps, identifying a need for caution in their application. 
Finally, we propose ``geometric engineering" as a new design paradigm, enabling the independent tuning of geometric responses without altering the energy dispersion.

\end{abstract}
\maketitle

\textit{Introduction.}
The quantum geometric tensor (QGT) whose symmetric part is the quantum metric and whose antisymmetric part is the Berry curvature~\cite{provost1980riemannian, ma2010abelian, berry1989quantum} has become central to contemporary condensed-matter physics~\cite{gao2025quantum, torma2022superconductivity, yu2025quantum, torma2023essay}. 
The Berry curvature underlies topological classification~\cite{nagaosa2010anomalous, Xiao_2010_RMP}, while the quantum metric has shown to influence a wide range of phenomena such as transport properties~\cite{oh2026mass,oh2025color,ezawa2024analytic,das2023intrinsic,gao2023quantum,wu2024quantum,mitscherling2022bound,huhtinen2023conductivity,ghosh2024probing,neupert2013measuring,oh2024thermoelectric,wang2023quantum,han2024room,han2026klein}, superconducting properties~\cite{peotta2015superfluidity,liang2017band,torma2022superconductivity,oh2025role}, magnetic properties~\cite{rhim2020quantum,jung2024quantum,oh2024revisiting, hwang2021geometric,oh2026magnetic,kitamura2024spin,gao2015geometrical, shimizu2026magnetic}, bulk-edge correspondence~\cite{oh2022bulk,kim2023general},
electron-phonon coupling~\cite{yu2024non}, and capacitance~\cite{komissarov2024quantum}.

In practice, the QGT is routinely calculated from tight-binding or Wannier-based models via $k$-derivatives of the occupied-band projector ($P$): 
\begin{align}
 Q_{\mu\nu}^{\partial}(\bm{k}) = \mathrm{Tr}\,[ \partial_\mu P(\bm{k}) (1-P(\bm{k})) \partial_\nu P(\bm{k}) ], \label{eq:QGT_conv}
\end{align}
where $\partial_\mu=\frac{\partial}{\partial k_\mu}$.
This definition is manifestly gauge-invariant within the occupied subspace, and it has become the standard tool for connecting microscopic models to geometric observables.
Yet, this definition hides an often-implicit and physically consequential choice: which Bloch convention is adopted when building the Bloch basis? 
To be concrete, let $\bm{R}$ denote the Bravais-lattice vector specifying {the position of} the unit cell, and let $\bm{x}_\alpha$ be the position of orbital or sublattice $\alpha$ measured from the origin of that unit cell.
One may then either omit the intra-unit-cell orbital positions from the Bloch phase, as in Convention A, $e^{i\bm{k}\cdot\bm{R}}$, or include them explicitly, as in Convention B, $e^{i\bm{k}\cdot(\bm{R}+\bm{x}_\alpha)}$.
This choice is frequently overlooked as a mere technical convenience, a view partially justified because the Brillouin zone (BZ) integral of the QGT's imaginary part yields a convention-independent Chern number for fully occupied bands~\cite{simon2020contrasting}.
This invariance, however, does not generalize to the QGT's real part. 
Even the imaginary part becomes convention-dependent for pointwise values or for the response of partially filled bands~\cite{simon2020contrasting}. 
Related basis- and embedding-sensitive issues have also emerged in modern orbital magnetism and orbital Hall response~\cite{sastges2026modern,lee2026anatomy}, warning that derivative-based geometric quantities may lead to incorrect results unless basis-state derivatives and nonlocal contributions are properly accounted for.
This issue has also begun to surface in flat-band physics, where the quantum-metric length is embedding dependent and proposed an embedding-independent flat-band length as the relevant intrinsic scale~\cite{lee2025embedding}.

In this paper, we demonstrate that within the standard framework, adopting Convention B is not a matter of choice but is uniquely mandated by the microscopic derivation of physical observables. When electromagnetic coupling is introduced via the Peierls substitution, the gauge field couples to the full coordinates. This derivation reveals that the physical current naturally accounts for intra-unit-cell orbital positions. While this information is intrinsic to the standard $k$-derivative in Convention B, it is erroneously omitted in Convention A. Consequently, many previous calculations that implicitly adopt Convention A without specifying orbital embedding are susceptible to leading-order physical errors.

To resolve this conceptual and practical failure, we introduce a convention-independent QGT ($Q^{\rm phys}$). The key is to replace the standard partial derivative $\partial_\mu$ with a covariant derivative, $D_\mu = i[\cdot,\hat{r}_\mu]$, which explicitly incorporates the full position operator $\hat{r}= \hat{R}+\hat{X}$. Here, $\hat{R}$ represents the unit-cell position operator associated with the Bravais-lattice coordinate $\bm R$, while $\hat{X}$ represents the intra-cell position operator associated with the orbital positions $\bm x_\alpha$.
When acting on the projector $P^{(A)}$, this derivative yields $D_\mu P^{(A)}=\partial_\mu P^{(A)}-i[\hat{X},P^{(A)}]$, while it reduces to the standard derivative $D_\mu P^{(B)}=\partial_\mu P^{(B)}$ for Convention B. This resulting $Q^{\rm phys}$ provides a unique, physical value regardless of the convention choice, perfectly resolving the ambiguities found in previous approaches.

In what follows, we first demonstrate the failure of the conventional partial-derivative definition QGT in Eq.~(\ref{eq:QGT_conv}) within Convention A by highlighting the unit-cell paradox within the Su-Schrieffer-Heeger (SSH) model, thereby establishing the necessity for a convention-independent formulation. We then derive a physical QGT that is uniquely mandated by the microscopic current via the Peierls substitution. Furthermore, we delineate the criteria for the validity of $k \cdot p$ effective theories, identifying a leading-order failure in systems with bond-ordered gaps. We conclude by proposing ``geometric engineering" as a design paradigm for metamaterials, where geometric responses can be tuned independently of spectral properties.

\textit{Emergence of QGT from physical observables.}
The QGT for the occupied bands is usually defined using the projector onto the occupied subspace, 
\begin{align}
    P(\bm k)=\sum_{n\in\mathrm{occ}}|u_{n \bm k}\rangle\langle u_{n \bm k}|,
\end{align}
where $|u_{n\bm k}\rangle$ denotes the cell-periodic Bloch eigenstate of a Bloch Hamiltonian $H(\bm k)$,
satisfying $H(\bm k)|u_{n\bm k}\rangle=\varepsilon_{n\bm k}|u_{n\bm k}\rangle$.
Here, $H(\bm k)$ refers to the Bloch Hamiltonian in a specified Bloch basis, whose convention dependence will be discussed below.
The conventional QGT is then given by Eq.~(\ref{eq:QGT_conv})~\cite{ma2010abelian}.
A primary virtue of this definition is its gauge invariance. Here, ``gauge" refers to the $\bm k$-dependent $U(N)$ unitary transformation within the occupied-band subspace, $|u_{n\bm k}\rangle \to \sum_m U_{mn}(\bm k) |u_{m\bm k}\rangle$ (for $n, m \in \text{occ}$). The QGT, being defined via the projector $P(\bm k)$ and a trace, is invariant under this $U(N)$ gauge transformation. This property has established the QGT as a physically meaningful quantity.

The QGT is not merely a mathematical abstraction of the Hilbert space bundle; it is a fundamental object that emerges from the microscopic derivation of physical observables.
For instance, the real part of the linear optical conductivity in two-band systems with $\omega>0$ is generally related to the quantum metric, $g_{\mu\nu} = \mathrm{Re}\,Q_{\mu\nu}$, by~\cite{ezawa2024analytic,oh2026mass} 
\begin{align}
\text{Re}[\sigma_{\mu\nu}(\omega)]= \pi\frac{e^2}{\hbar}\int_{\mathrm{BZ}}\frac{d^dk}{(2\pi)^d} \Delta(\bm k)g_{\mu\nu}(\bm k)\delta(\hbar\omega-\Delta(\bm k)),\label{eq:optkubo}
\end{align}
where $\Delta(\bm{k})$ indicates the energy difference between conduction and valence bands.
Another example is the spread of a Wannier function $\Omega$. This can be decomposed into a gauge-invariant part and a gauge-dependent part. The gauge-invariant part $\Omega_I$ represents the intrinsic limit of how localized Wannier orbitals can be. It has been rigorously shown that $\Omega_I$ is the BZ integral of the trace of the quantum metric~\cite{marzari1997maximally,yu2025quantum} 
\begin{align}
    \Omega_I = \int_{\text{BZ}} \frac{d^d k}{(2\pi)^d} \mathrm{Tr}\,g_{\mu\nu}(\bm{k}). \label{eq:Wannier}
\end{align}

While the emergence of $g_{\mu\nu}$ in these observables is well-documented, a subtle but critical point is often overlooked: these relations are valid only when the momentum derivatives $\partial_{\bm k}$ are consistent with the full position operator $\hat{r}$, i.e. only in Convention B.
In the context of optical conductivity, according to the fundamental principle of minimal coupling (or Peierls substitution), the physical current operator is defined by the commutator of the Hamiltonian with the full position operator $\hat{r}$, rather than a partial momentum derivative of the Hamiltonian $\partial_{k_i} H$.
Namely, the position of an electron in a crystal should necessarily include its coordinate $\hat{X}$ within the unit cell. 
Therefore, the definition of a Wannier function's spread in real space necessitates a Bloch basis in Convention B that explicitly incorporates the intra-cell orbital, i.e. sublattice coordinates $x_\alpha$. Consequently, Convention A, which treats all orbitals as if they sit at the unit-cell origin ($x_\alpha = 0$) during the $k$-differentiation process, introduces a fundamental discrepancy between the computational geometry with Eq.~(\ref{eq:QGT_conv}) and the actual physical observables.

\textit{Ambiguity and Failure of the QGT in Convention A.}
We focus on the convention dependence arising from the definition of the Bloch transform itself. While the QGT is $U(N)$ gauge-invariant, we demonstrate that it is not invariant under the choice of the Bloch basis representation.



The Bloch basis in Convention A (the cell-periodic representation), where intra-cell orbital positions are omitted, is defined as 
\begin{align}|u^{(A)}_{\bm k\alpha}\rangle=\frac{1}{\sqrt{N}}\sum_{\bm R} e^{-i \bm k \cdot \bm R}|\phi_{\bm R\alpha}\rangle,\end{align}
where $\phi_{\bm{R} \alpha}$ denotes the atomic orbital at unit cell $\bm{R}$ with sublattice or orbital index $\alpha$, and $N$ is the number of unit cells.
It is widely observed that Convention A is adopted in the literature~\cite{jankowski2025excitonic,ezawa2024analytic,chen2024enhancing,zeng2024quantum,xu2019nonlinear,terada2025multitunneling}.
On the other hand, Convention B (the orbital position embedded representation) is
\begin{align}|u^{(B)}_{\bm k\alpha}\rangle=\frac{1}{\sqrt{N}}\sum_{\bm R} e^{-i \bm k\cdot (\bm R+\bm x_\alpha)}|\phi_{\bm R\alpha}\rangle.
\end{align}
The two bases are connected by the diagonal, $k$-dependent unitary transformation $S(\bm k)=\exp(i \bm k\cdot\hat X)=\mathrm{diag}\big(e^{i \bm k\cdot \bm x_\alpha}\big)$, such that $|u^{(A)}_{\bm k\alpha}\rangle = e^{i \bm k\cdot \bm x_\alpha}\,|u^{(B)}_{\bm k\alpha}\rangle$. Consequently, the projectors and Hamiltonians are related by $P^{(A)}(\bm k)=S(k)\,P^{(B)}(\bm k)\,S^\dagger(\bm k)$, $H^{(A)}(\bm k)=S(\bm k)\,H^{(B)}(\bm k)\,S^\dagger(\bm k)$.
The QGT definition in Eq.~(\ref{eq:QGT_conv}) involves the $\partial_\mu$ operator, which makes this $\bm k$-dependence problematic. Applying the derivative to $P^{(A)}$ yields terms from the derivative of $S(\bm k)$ compared to Convention B.
As a result, the QGT as defined in Eq.~(\ref{eq:QGT_conv}) is not invariant under this transformation: $Q_{\mu\nu}^{\partial,(A)}(\bm k) \neq Q_{\mu\nu}^{\partial,(B)}(\bm k)$.

This convention dependence acts in an asymmetric fashion on the imaginary and real parts of the QGT.
While the Berry curvature $F_{\mu\nu}(\bm k)$ differs between conventions, the difference is a total derivative. Therefore, the Chern number, i.e. an integral of the Berry curvature, remains convention-independent for fully occupied bands~\cite{simon2020contrasting}.
In contrast, the real part $g_{\mu\nu}$ enjoys no such protection. Not only do the local values differ, but their BZ-integral also depends on the chosen convention. 

Physical observables such as Eqs.~(\ref{eq:optkubo}) and (\ref{eq:Wannier}) cannot depend on an arbitrary choice between Conventions A and B. From the derivations of Eqs.~(\ref{eq:optkubo}) and (\ref{eq:Wannier}), Convention B should be adopted here. It is natural because the Convention A is problematic in a sense that its $k$-derivative explicitly ignores the intra-unit cell coordinates $\bm{x}_\alpha$, treating all orbitals as if they are collapsed onto the unit-cell origin during the calculation of geometric phases. As we will show in the next section using the SSH model, this omission leads to a ``unit-cell paradox" where shifting the arbitrary boundary of a unit-cell changes the calculated physical response—a result that is clearly unphysical.

\begin{figure}[t]
\includegraphics[width=80mm]{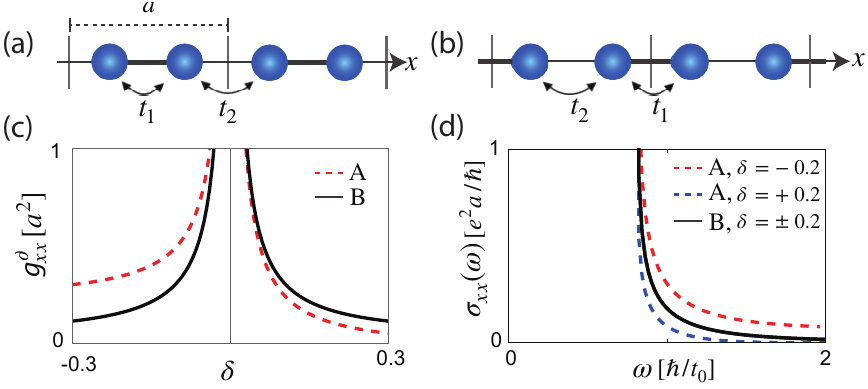} 
\caption{
The unit-cell paradox in the SSH model.
(a) Lattice structure of the SSH model, defining $t_1$ as the intra-cell hopping (bold line) and $t_2$ as the inter-cell hopping (thin line), with lattice constant $a$. (b) The identical physical chain with a shifted unit cell partition, equivalent to a $t_1 \leftrightarrow t_2$ exchange. 
(c) The BZ-integrated quantum metric (total Wannier spread) $\mathcal{G}_{xx}$ as a function of the hopping asymmetry parameter $\delta$ that quantifies the dimerization. Here, we set $t_1 = t_0(1+\delta)$ and $t_2 = t_0(1-\delta)$. The dashed line corresponds to $\mathcal{G}_{xx}^{\partial,(A)}$ by Convention A, while the solid line corresponds to $\mathcal{G}_{xx}^{\rm phys}$ by Convention B.
(d) Optical conductivity spectrum $\sigma_{xx}(\omega)$ for $\delta = \pm 0.2$ (solid line), whereas the dashed lines are obtained by inserting $g_{xx}^{\partial,(A)}$ into Eq.~(\ref{eq:optkubo}).
\label{fig1}}
\end{figure}

\textit{The unit-Cell Paradox in the SSH Model.}
To demonstrate that the convention dependence of the QGT in Eq.~(\ref{eq:QGT_conv}) can lead to unphysical results, we apply the formula of $Q_{\mu\nu}^\partial$ in Eq.~(\ref{eq:QGT_conv}) to the SSH model~\cite{su1979solitons}. The model is a 1D tight-binding chain along the $x$-axis with lattice constant $a$ and alternating hopping amplitudes: $t_1$ (intra-cell) and $t_2$ (inter-cell), as shown in Fig.~1(a,b). 

A basic requirement is that bulk observables must be independent of arbitrary human choices, such as how the lattice is partitioned into unit cells.
Consider two infinite SSH chains described by $(t_1, t_2)$ and $(t_2, t_1)$. These are physically the same bulk system: the latter is obtained simply by shifting the unit-cell origin by half a lattice constant, so that the old intra-cell bond becomes the new inter-cell bond, and vice versa. Therefore, any bulk physical observable $f$ must satisfy $f(t_1, t_2) = f(t_2, t_1)$.
To test this, we parameterize the hoppings as $t_1 = t_0(1+\delta)$ and $t_2 = t_0(1-\delta)$, where $\delta$ is the hopping asymmetry parameter. The exchange $t_1 \leftrightarrow t_2$ corresponds to $\delta \leftrightarrow -\delta$.

We first evaluate $Q_{\mu\nu}^{\partial,(A)}$ (see Supplemental Materials for details), for which $H^{(A)}(k) = (t_1 + t_2 \cos (ka))\sigma_x + t_2 \sin (ka)\sigma_y$, where $\sigma_i$ are Pauli matrices.
Our first test is the BZ-integrated quantum metric
$\mathcal{G}^{\partial,(A)}_{xx} = \int \frac{adk}{2\pi} g^{\partial,(A)}_{xx}(k)$.
In one dimension, the gauge-dependent part of the Wannier spread can be removed in the maximally localized gauge, so the physical spread is determined by the appropriate physical geometric tensor~\cite{marzari1997maximally,yu2025quantum}. The point of Fig.~1(c), however, is that if one inserts the metric from Convention A into the usual spread formula, one obtains an unphysical result: $\mathcal{G}^{\partial,(A)}_{xx}$ is manifestly asymmetric under $\delta \leftrightarrow -\delta$, even though the underlying bulk chain is unchanged.

Our second test is the interband optical conductivity by inserting $g^{\partial,(A)}_{xx}(k)$ into Eq.~(\ref{eq:optkubo}). Since the band gap $\Delta(k)$ is symmetric under $\delta \leftrightarrow -\delta$, any asymmetry in the resulting spectrum must originate from $g^{\partial,(A)}_{xx}(k)$. As shown in Fig.~\ref{fig1}(d), the spectra for $\delta = \pm 0.2$ calculated via Convention A (dashed lines) are distinct, which is unphysical.

The problem is resolved by employing the $Q_{\mu\nu}^\partial$ within Convention B, where the orbital positions are explicitly embedded in the Bloch phase. 
Here, we choose the relative distance between the two sublattices as $a/2$.
As shown by the solid lines in Figs.~\ref{fig1}(c) and \ref{fig1}(d), the physical calculation perfectly restores the $t_1 \leftrightarrow t_2$ symmetry. The Wannier spread becomes an even function of $\delta$, and the optical conductivity spectra for $\delta$ and $-\delta$ become identical. These numerical calculations confirm that the QGT calculated by the
ordinary derivative in Convention A is physically incomplete and therefore cannot in general be used directly in Eqs.~(\ref{eq:optkubo}) and (\ref{eq:Wannier}), because it omits the intra-cell position information encoded in $\hat{X}$.

\textit{Physical Current and convention-independent QGT.}
Since the QGT is fundamentally linked to measurable physical observables, it must be defined in a convention-independent manner. We introduce the convention-independent QGT that is uniquely mandated by the microscopic derivation of the current operator via the Peierls substitution.

The physical current operator $J$ is defined by coupling the Hamiltonian to an external vector potential $\bm{A}$ and taking the functional derivative:
$$J_\mu = -\frac{\delta H(\bm{A})}{\delta A_\mu} \bigg|_{\bm{A}=0}$$
When we apply this to a tight-binding Hamiltonian, the Peierls phase $e^{i\frac{e}{\hbar} \bm{A} \cdot (\bm{r}_j - \bm{r}_i)}$ must account for the full coordinates of the orbitals, $\bm{r}_i = \bm{R}_i + \bm{x}_\alpha$.

If we calculate the current operator using the Hamiltonian in Convention B ($H^{(B)}$), the result is 
\begin{align}
    J^{(B)}_\mu(\bm k) \equiv -\frac{e}{\hbar}\frac{\partial H^{(B)}(\bm k)}{\partial k_\mu}
\end{align}
This shows that the current in Convention B is given simply by the k-derivative of the Hamiltonian. Because Convention B embeds the orbital positions $x_\alpha$ directly into the Bloch phase $e^{i \bm{k} \cdot (\bm{R} + \bm{x}_\alpha)}$, the $k$-derivative automatically extracts the intra-cell coordinates. 
Thus, the current within Convention B is naturally physical.

By contrast, in Convention A, a correction term involving the intra-cell position operator $\hat{X}$ appears
\begin{align}J^{(A)}_\mu(\bm k) &=S(\bm{k})J_\mu^{(B)}(\bm{k})S^\dagger(\bm k) \nonumber \\
&= \underbrace{-\frac{e}{\hbar}\frac{\partial H^{(A)}(\bm{k})}{\partial k_\mu}}_{J^{\partial, (A)}} + \underbrace{\frac{e}{\hbar}i\big[\hat{X}_\mu,H^{(A)}(\bm{k})\big]}_{J^{\rm corr}} .\label{eq:Jtrue_A}
\end{align}
The term $J^{\rm corr}$ represents the current contribution arising from the electronic motion within the unit cell. The failure of the ordinary projector-derivative QGT in Convention A stems precisely from the omission of this term; the operator $J^{\rm \partial, (A)}$ only accounts for hopping between unit cells.
Indeed, the necessity of such correction terms has been previously recognized in the context of ensuring gauge invariance in nonlinear optical responses~\cite{ventura2017gauge}. 
However, while previous works utilized this correction primarily to rectify response functions, we demonstrate that it points to a more fundamental issue: the QGT itself must be defined covariantly to consistently describe the geometric properties of the band structure.

To provide a convention-independent framework that yields unique physical values, we propose replacing the standard partial derivative $\partial_\mu$ with a covariant derivative, $D_\mu$. We define this operator such that it incorporates the full position operator $\hat{r} = \hat{R} + \hat{X}$:
\begin{align}
D_\mu P \equiv i[P, \hat{r}_\mu].
\end{align}
In the representation of Convention A, this derivative acts on the projector $P^{(A)}$ as $D_\mu P^{(A)} = \partial_\mu P^{(A)} - i[\hat{X}_\mu, P^{(A)}]$.
In contrast, for Convention B, it reduces to the standard derivative $D_\mu P^{(B)} = \partial_\mu P^{(B)}$.
Using this covariant derivative, we define the physical QGT ($Q^{\rm phys}$) as
\begin{align} Q^{\rm phys}_{\mu\nu}(\bm k)= \mathrm{Tr}\big[D_\mu P(\bm k)(1-P(\bm k))D_\nu P(\bm k)\big]
\label{eq:QGT_phys}.\end{align}
This formulation ensures that the resulting geometric tensor is invariant under the choice of Bloch convention.
Note that this formulation aligns with the modern theory of polarization, where the position operator in periodic systems is rigorously defined via the quantum mechanical phase of the flux~\cite{resta1998quantum, resta1994macroscopic}. 

\textit{Implications for Low-Energy Effective Theories.}
Up to this point, we have established within tight-binding lattice models that the physical QGT must be defined via a covariant derivative that includes the orbital position operator $\hat{X}$. However, a vast amount of contemporary research—particularly in topological materials and twistronics—relies not on full lattice models, but on low-energy effective Hamiltonians ($k \cdot p$ theory) derived near high-symmetry points in the BZ.
A question naturally arises: does the same issue persist in $k\cdot p$ theories?
The answer is yes, but it depends crucially on the physical origin of the energy gap.

To discuss this issue, we consider a low-energy effective Hamiltonian $H_{\mathrm{eff}}(\bm q)$ near a high-symmetry point $\bm k_0$, where $\bm q$ denotes the small momentum measured from $\bm k_0$. 
The projector $P_{\mathrm{eff}}$ onto the low-energy subspace at $\bm k_0$ defines an effective embedding operator $X_{\mathrm{eff},\mu} = P_{\mathrm{eff}}\hat X_\mu P_{\mathrm{eff}}$. 
Physically, $X_{\mathrm{eff}}$ encodes the relative intra-cell positions of the effective orbitals or sublattices retained in the low-energy theory. 
Applying Eq.~(\ref{eq:Jtrue_A}) to the
projected theory, the physical current takes the form
$J_{\mathrm{eff},\mu}(\bm q)= -\frac{e}{\hbar} \partial_{q_\mu}H_{\mathrm{eff}}(\bm q) +  J_{\mathrm{eff},\mu}^{\mathrm{corr}}(\bm q)$, where
\begin{align}
J_{\mathrm{eff},\mu}^{\mathrm{corr}}(\bm q)
&=
\frac{ie}{\hbar}
\left[
X_{\mathrm{eff},\mu},
H_{\mathrm{eff}}(\bm q)
\right].
\label{eq:effective_embedding_current}
\end{align}
Thus, the geometric and optical response of the effective theory is,
in general, determined not by \(H_{\mathrm{eff}}\) alone, but by the
pair \((H_{\mathrm{eff}},X_{\mathrm{eff}})\).

From this perspective, low-energy effective theories can be broadly divided into gapless and gapped classes. For gapless effective theories, one typically has $H_{\mathrm{eff}}(\bm 0)=0$. 
Assuming an analytic low-energy expansion, $H_{\mathrm{eff}}(\bm q)=\sum_\nu q_\nu V_\nu+O(q^2)$, Eq.~(\ref{eq:effective_embedding_current}) gives $ J_{\mathrm{eff},\mu}^{\mathrm{corr}}(\bm q)=O(q)$, whereas the ordinary $k\cdot p$ current $-e\partial_{q_\mu}H_{\mathrm{eff}}/\hbar$ begins at $O(q^0)$.
Orbital embedding therefore does not modify the zeroth-order current vertex or the associated leading low-energy geometric response.
The standard $k\cdot p$ description remains reliable at this order,
although embedding effects can appear in subleading powers of $\bm q$.

For a gapped theory, by contrast, $H_{\mathrm{eff}}(\bm 0)$ contains a matrix-valued gap term, and the embedding correction at the expansion point is $J_{\mathrm{eff},\mu}^{\mathrm{corr}}(\bm 0)$.
This correction is generically of zeroth order in \(\bm q\) and therefore contributes at the same order as the conventional $k\cdot p$ current vertex. Consequently, the response of a gapped effective theory is not, in general, fixed by $H_{\mathrm{eff}}$ alone.
Within this gapped class, simple minimal models illustrate two representative possibilities. 
The first case is the commuting case defined by $[H_{\mathrm{eff}}(0),X_{\mathrm{eff}}]=0$.
This is exemplified by an on-site mass term such as $H_{\mathrm{eff}}(0)=M\sigma_z$, when $X_{\mathrm{eff},\mu}$ is diagonal in the same orbital basis. 
The standard $k\cdot p$ current then remains valid at leading order, although embedding corrections may still enter at higher orders in $\bm q$.
In the noncommuting case, $[H_{\mathrm{eff}}(0),X_{\mathrm{eff},\mu}]\neq 0$, orbital embedding modifies the current already at zeroth order.
This situation is realized for SSH~\cite{su1979solitons} or Kekul\'e~\cite{hou2007electron} type bond-order masses, where the gap is generated by inter-sublattice or inter-orbital hopping terms rather than by an on-site potential difference. 

\begin{figure}[t]
\includegraphics[width=80mm]{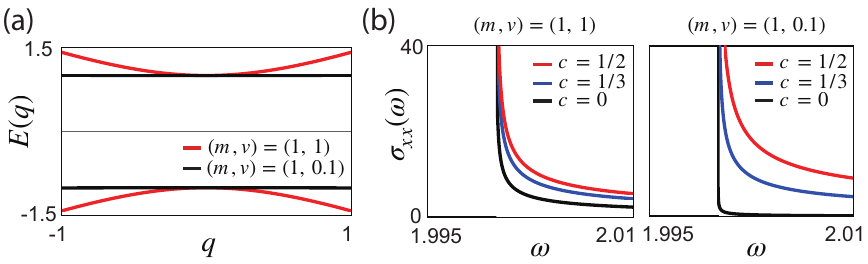} 
\caption{
Impact of effective orbital embedding in the one-dimensional massive Dirac model. (a) Energy dispersion $E(q)$ for a dispersive band ($v=1$, red line) and a flatter band ($v=0.1$, black line) at fixed mass $m=1$.
(b) Interband optical conductivity $\sigma_{xx}(\omega)$ for different values of the effective embedding parameter \(c\) and different choices of \((m,v)\).
\label{fig2}
}
\end{figure}

To illustrate the noncommuting case, consider the one-dimensional massive Dirac Hamiltonian
\begin{align}
H_{\mathrm{1D}}(q)=vq\,\sigma_x+m\,\sigma_y.
\end{align}
Although this Hamiltonian fixes the low-energy dispersion, it does not uniquely fix the optical conductivity. 
For a two-orbital effective model, let $c$ denote the relative position between the two sublattice/orbital degrees of freedom. Equivalently, one may choose $X_{\mathrm{eff}}=-(c/2)\sigma_z$. The parameter $c$ plays a role of the embedding parameter.
As shown in Fig.~\ref{fig2}, the same continuum Hamiltonian can yield markedly different optical responses depending on $c$, demonstrating that, in gapped effective theories, $H_{\mathrm{eff}}(q)$ alone is insufficient to determine the physical geometric response.

At the band edge, $\omega\to 2|m|^+$, the ratio between the conductivity and the embedding-blind result is
\begin{align}
\lim_{\omega\to2|m|^+}
\frac{\mathrm{Re}[\sigma^{\mathrm{phys}}(\omega)]}
{\mathrm{Re}[\sigma^{\partial}(\omega)]}
=
\left(1+\frac{cm}{v}\right)^2.
\end{align}
Here, $\sigma^{\mathrm{phys}}$ is computed using the current including $X_{\mathrm{eff}}$, whereas $\sigma^\partial$ is computed using only the ordinary derivative $\partial_q H_{\mathrm{1D}}$. 
In other words, the former is computed using the projected full physical current $J_q^{(A)}$ in Eq.~(\ref{eq:Jtrue_A}), or equivalently $J_q^{(B)}$ in Convention B, while the latter corresponds to retaining only the embedding-blind component $J_q^{\partial,(A)}$ of that current.
This shows that the discrepancy is strongly amplified as the band becomes flatter ($v\to 0$). In flat or nearly flat bands, orbital embedding is therefore not a small correction, but an essential part of the low-energy data relevant for geometric and optical response.

\textit{Geometric Engineering.}
By regarding the Hamiltonian $H(\bm k)$ and the orbital embedding $\hat{X}$ to be independent inputs, one opens a new design paradigm: Geometric Engineering. 
Since the energy dispersion is determined solely by hopping parameters, while geometric observables are sensitive to both hopping parameters and $\hat{X}$, it is theoretically possible to construct systems with identical energy spectra but distinct quantum geometries. 
This is achieved by tuning the spatial embedding of orbitals while maintaining constant effective hopping parameters.

Classical metamaterials, including mechanical, acoustic, and photonic lattices, provide an ideal platform for this concept, as coupling strengths and intra-cell coordinates can be controlled independently. 
As a concrete example, we consider a mechanical analogue of the SSH model: a one-dimensional spring-mass model with identical masses at two sites per unit cell, connected by alternating springs with spring constants $K_1$ and $K_2$, as shown in Fig.~3.
Two such mechanical SSH chains can be engineered to share identical vibrational bands by keeping the spring constants and lattice constant fixed, while changing the relative intra-cell spacing of the masses.
%
Although these systems share the same dynamical matrix $D^{(A)}(k)$ and hence the same spectrum, their physical QGT differs because the embedding $\hat X$ is different. 
Consequently, they can exhibit distinct geometric signatures, such as different Wannier spreads, spatial extent of localized modes, and response functions. This illustrates how geometric engineering, rather than spectral engineering alone, can serve as an independent design principle for functional metamaterials.

\begin{figure}[t]
\includegraphics[width=80mm]{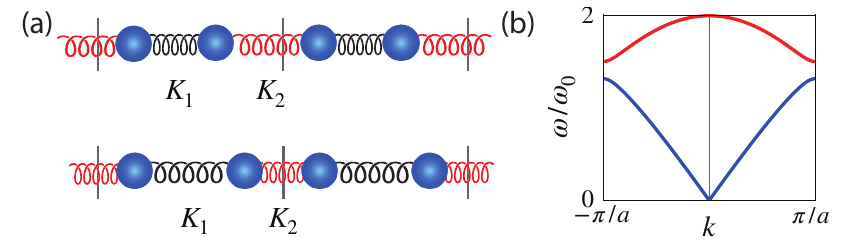} 
\caption{
Spectral invariance under geometric engineering. (a) Mechanical SSH chains with distinct spacings but identical force constants $K_j$ and the lattice constant $a$. (b) Vibrational bands $\omega/\omega_0$. The bands remain invariant under geometric tuning, as the spectra depend on $K_j$, $a$ and mass $m$. In contrast, the physical QGT is sensitive to the intra-cell spacing, providing a platform for geometric engineering (see Appendix). Parameter $\omega_0$ is defined as $\sqrt{(K_1+K_2)/(2m)}$. Here, we used $K_2/K_1=1.4$.
\label{fig3}}
\end{figure}

\textit{Conclusion.}
This study demonstrates that the QGT calculations are critically sensitive to unit-cell conventions, leading to a fundamental unit-cell paradox. Using the SSH model, we proved that the conventional $k$-derivative approach within Convention A yields unphysical, non-invariant results for observables.
We resolved this inconsistency by introducing a physical QGT defined via a covariant derivative, $D_\mu = i[\cdot, \hat{r}_\mu]$. This framework, uniquely mandated by the microscopic current derived from Peierls substitution, explicitly incorporates the orbital position operator $\hat{X}$ to ensure convention-independent results.
Notably, our work highlights a need for caution when applying not only lattice models but also $k \cdot p$ low-energy effective theories: the standard approximation fails to capture the leading-order geometric contributions whenever the energy gap originates from bond terms, a discrepancy that can be dramatically amplified in the flat-band limit.

Furthermore, the independence of $H(k)$ and $\hat{X}$ establishes a new paradigm of geometric engineering. By decoupling spectral properties from geometric responses, we can independently tune physical observables, such as Wannier localization and optical absorption, without altering the energy dispersion. This provides a powerful degree of freedom for designing functional quantum materials and classical metamaterials, where performance is governed by the spatial embedding of the Hilbert space rather than spectral engineering alone.

\begin{acknowledgments}
The authors thank J. Rhim and D. Go for useful discussions.
We acknowledge the support by Japan Society for
the Promotion of Science (JSPS), KAKENHI Grant No. JP25KF0186, NO. JP22K18687, No. JP22H00108, and No. JP24H02231.
\end{acknowledgments}

\bibliography{ref.bib}

\end{document}